\documentclass[useAMS,usenatbib]{mn2e}

\usepackage{graphicx}
\usepackage{epstopdf}
\usepackage{amsmath}

\usepackage{times}

\title[]{Hydrodynamical wind on vertically self-gravitating ADAFs
in the presence of toroidal magnetic field}

\author[Maryam Ghasemnezhad and Shahram Abbassi ]{
Maryam Ghasemnezhad $^{1}$\thanks{E-mail:m$_{-}$ghasemnezhad2005@yahoo.com} and Shahram Abbassi $^{2,3}$ \thanks{E-mail:abbassi@um.ac.ir} \\
$^{1}$ Faculty of physics, Shahid Bahonar University of Kerman, Kerman, Iran\\
$^{2}$ Department of Physics, School of Sciences, Ferdowsi University of Mashhad, Mashhad, 91775-1436, Iran\\
$^{3}$ School of Astronomy, Institute for Research in Fundamental Sciences (IPM), Tehran, 19395-5531, Iran}

\date{}
\begin{document}
\pagerange{\pageref{firstpage}--\pageref{lastpage}} \pubyear{2015}\maketitle \label{firstpage}
\maketitle \label{firstpage}
\begin{abstract}
 We present the effect of a hydrodynamical wind on the structure and the surface temperature of
 a vertically self-gravitating magnetized ADAFs using  self-similar solutions. Also a model for an axisymmetric, steady-state,
 vertically self-gravitating hot accretion flow threaded by a toroidal magnetic field has been formulated.
 The model is based on $\alpha-$prescription for turbulence viscosity.
 It is found that the thickness and radial velocity of the disc are reduced significantly as wind gets stronger.
 In particular, the solutions indicated that the wind and advection have the same effects on the structure of the disc.
 We also find that in the optically thin ADAF becomes hotter by including the wind parameter and the self-gravity parameter.

\end{abstract}


\section{INTRODUCTION}
The accretion of matter into a compact object is a classical
problem in modern astrophysics and is an important tool for
understanding energetic phenomena including active galactic
nuclei, ultra luminous X-ray sources and galactic jets. The
standard theory of astrophysical accretion disc was formulated
over thirty years ago (Pringle \& Rees 1972; Novikov \& Thorne
1973; Shakura \& Sunyaev 1973, Kato et al. 2008). An accretion
disc is a structure formed by diffuse material in spiral motion
around a massive central body by losing their initial angular
momentum and some of the gravitational energy which liberate and
convert into radiation.

The accretion disc theory has been developed rapidly during the
past three decades. Since then a large body of observational data
has been accumulated, however, required some other types of
models distinct from the classical picture. Advective cooling is
one of the most important mechanism which are not considered in
the energy equations in the standard model. In another point of
view, if the gas density is low, the gas maybe unable to radiate
energy at a rate that balances viscous heating. In this case, the
heat generated by viscosity will be ÒadvectedÓ inwards with the
flow instead of being radiated. The disc becomes hot, hence
geometrically thick, low density, and radiatively inefficient.
Such ÒAdvection Dominated Accretion FlowsÓ (ADAFs) were
introduced by Narayan \& Yi (1994). In comparison to
thin accretion discs, advection dominated flows have quite
different structure, lower bolometric efficiency, and very
different SED (spectral energy distributions), with emission over
a wider range of wavelengths. It has been argued (Yuan \& Narayan
(2014), a comprehensive review and reference there in), based on
a combination of theoretical models and observational studies of
stellar mass black holes, that the inner regions of accretion
discs are replaced by ADAFs when the accretion rate decreases
below $ \sim 0.01\dot{M}_\mathrm{Edd}$. The dynamical and radiative
properties of the ADAFs have been intensely studied during the
past several years. Consequently, the model has been applied to
astrophysical black hole systems, such as supermassive black hole
in our galactic center, Sagittarius $A^*$, low-luminosity AGNs
(LLAGNs), black hole binaries in their hard and quiescence stats.
Although ADAF works well for Sgr A* and some LLAGNs, many details
of ADAF need to be investigated (e.g., the dynamical role of
magnetic field; the influence of outflow; etc). In order to
deepen our understanding of the accretion process modeling to
more sources is required.

The importance and presence of magnetic fields in the accretion
discs is generally accepted. The dynamical importance of magnetic
field widely recognized in angular-momentum transport, the
formation of jets-outflow, and the interactions between holes and
discs, etc. Several attempt have been done for studying
magnetized accretion flows analytically (Kaburaki (2000), Akizuki
\& Fukue (2006), Abbassi et al. (2008, 2010), Ghasemnezhad et
al.(2012)). More or less, they confirmed that in the presence of
a magnetic field, the structure and dynamics of the flow will
change considerably. So is appearently  essential to develop
ADAFs model for a fully magnetized case. On the other hand,
observations and theoretical arguments show that
hot accretion flows are associated with outflow ( Blandford \& Begelman 1999).  A physically more satisfactory approach had been proposed by (Akizuki \& Fukue (2006), Abbassi et al 2008, 2010, Ghasemnezhad 2013) by adding outflow and wind effects on the radial structure of ADAFs.\\

In the pioneer ADAFs paper the effect of self-gravity in the
vertical (or even in radial direction) is completely neglected
for simplicity and assumed that the disc is supported in the
vertical direction only by the thermal pressure. Mosallanezhad et
al. (2012) studied the vertical self-gravitating ADAFs in the
presence of toroidal magnetic field. Their solutions showed the
thickness of the disc modified significantly when self-gravity
becomes stronger. In this paper we develop Mosallanezhad et al.
(2012) solutions by considering the effect of wind in the MHD
equations. The hypothesis of the model and relevant equations are
devenoped Sec. 2. Self-similar solutions are presented in section
3. We show the result in section 4 and finally we present the
summary and conclusion in section 5.

\section{The Basic Equations}
Our goal here is to study the effect of wind on the magnetized
ADAF in the presence of the vertical self-gravity of the disc. We
ignored the self-gravity in radial direction. We used vertically
integrated MHD equations in cylindrical coordinates $(r, \varphi,
z)$ for steady state and axi-symmetric ($\frac{\partial}{\partial
\phi}=\frac{\partial}{\partial t}=0$) hot accretion disc. we
suppose that all flow variables are only a function of $r$
(radial direction). We ignore the relativistic effects and we use
Newtonian gravity in the radial direction.\ We suppose that the
gaseous disc is rotating around a compact object of mass
$M_{\star}$. By adopting $\alpha$ -prescription for viscosity of
rotating gas in accretion flow. We consider the magnetic field
has just toroidal component.

The equation of continuity gives:

\begin{equation}
\frac{\partial}{\partial r}(r \Sigma V_r )+\frac{1}{2\pi}\frac{\partial \dot{M}_\mathrm{w}}{\partial r}=0
\end{equation}
where $V_{r}$ is the accretion velocity ($V_{r}<0$) and
$\Sigma=2\rho H$ is the surface density at a cylindrical radius
$r$. $H$ is the disc half-thickness and $\rho$ is the density.
Mass-loss rate by wind is showed by $\dot{M}_\mathrm{w}$. So

\begin{equation}
\dot{M}_\mathrm{w}=\int 4\pi r'\dot{m}_\mathrm{w}(r')dr',
\end{equation}
where $\dot{m}_\mathrm{w}(r)$ is the mass-loss per unit area
from each disc face. On the other hand, we can rewrite the
continuity equation as:
\begin{equation}
\frac{1}{r}\frac{\partial}{\partial r}(r\Sigma V_r )=2 \dot{\rho}H
\end{equation}

where $\dot{\rho}$ is the mass loss rate per unit volume.  The
equation of motion in the radial direction is:

\begin{displaymath}
V_r \frac{\partial V_r}{\partial
r}=\frac{V_{\varphi}^{2}}{r}-\frac{GM_{\star}}{r^2}-\frac{1}{\Sigma}\frac{d}{dr}(\Sigma
c_\mathrm{s}^{2})
\end{displaymath}

\begin{equation}
-\frac{c_\mathrm{A}^{2}}{r}-\frac{1}{2\Sigma}\frac{d}{dr}(\Sigma c_\mathrm{A}^{2})
\end{equation}

 where $V_{\varphi}, G, c_\mathrm{s}$, and $c_\mathrm{A}$ are the rotational velocity of
 the flow, the gravitational constant, sound speed and Alfven velocity of the
 gas, respectively. The sound speed and the Alfven velocity are
 defined as $c_\mathrm{s}^{2}=
 \frac{p_\mathrm{gas}}{\rho}$ and
 $c_\mathrm{A}^{2}=\frac{B_{\varphi}^{2}}{4\pi\rho}=\frac{2p_\mathrm{mag}}{\rho}$,
 where $B_{\varphi}$, $p_\mathrm{gas}$ and $p_\mathrm{mag}$ are the toroidal component of magnetic field, the gas and magnetic pressure
 respectively.

By integrating along $z$ of the azimuthal equation of motion
gives.
\begin{equation}
r\Sigma V_r \frac{d}{dr}(rV_\varphi)=\frac{d}{dr}(r^{3}\nu
\Sigma\frac{d\Omega}{dr})-\frac{\Omega(lr)^{2}}{2\pi}\frac{d\dot{M}_\mathrm{w}}{dr}
\end{equation}
where $\nu$ is the kinematic viscosity coefficient.
$\alpha$-prescription (Shakura \& Sunyaev 1973) for viscosity was
assumed as:
\begin{equation}
 \nu=\alpha c_\mathrm{s}H
 \end{equation}
where $\alpha$ is a constant less than unity.
$\Omega(=\frac{V_{\varphi}}{r})$ is the angular speed. To
write the angular momentum equation, we have considered the role
of wind in transferring the angular momentum. It is assumed that
the wind material moving along a steam line originating at radius
$r$ in the disc and co-rotate with the disc out to a
radial distance $lr$. The wind material ejected at radius $r$ on
the disc and carries away specific angular momentum
$(lr)^{2}\Omega$, where $\Omega$ related to a radial distance
$lr$. Knigge (1999) define the $l$ parameter as the length of the
rotational lever arm that allows us to have many types of accretion
disc winds models. The parameter $l=0$ corresponds to a
non-rotating wind and the angular momentum is not extracted by
the wind and the disc losses only mass because of the wind while
$l\neq1$ represents outflowing materials that carry away the angular
momentum (Knigge 1999, Abbassi et al. 2013).

By integrating along $z$ of the hydrostatic balance, we have:
\begin{equation}
H=\frac{c_\mathrm{s}^{2}(1+\beta)}{2\pi G \Sigma}
\end{equation}
where
$\beta=\frac{p_\mathrm{mag}}{p_\mathrm{gas}}=\frac{1}{2}(\frac{c_\mathrm{A}}{c_\mathrm{s}})^{2}$
indicates the important of magnetic field pressure compared to
gas pressure. We will study the dynamical properties of the disc
for different values of $\beta$.\\
Here we consider only self-gravity in the vertical direction, and
assume that in the radial direction centrifugal forces are
balanced by gravity from a central mass (Keplerian
approximation). One can estimate the importance of self-gravity
of the disc by comparing the contributions to the local
gravitational acceleration in the vertical direction by both the
central object and the disc itself. Here after, we will often
refer to such accretion discs, in which self-gravity is important
only in the vertical direction, as Keplerian self-gravitating
(KSG) discs (Duschl et al 2000).  The vertical
gravitation due to the disc's self-gravity at the disc surface is
given by ($2\pi G \Sigma$), and due to the central object is given by
$\frac{GM_{\star}H}{r^{3}}$. Thus, self-gravity of the disc in
vertical direction is dominated if (Mosallanezhad et al. 2012):

\begin{equation}
2\pi G \Sigma > \frac{G M_{\star}H}{r^{3}}
\end{equation}
\begin{equation}
  \frac{M_\mathrm{d}}{M_{\star}}> \frac{1}{2}\frac{H}{r}
\end{equation}
where $M_\mathrm{d}(r)=\pi r^{2}\Sigma$ is the mass enclosed in the disc
within a radius $r$.\\
For increasing disc masses self-gravity first becomes important
in the vertical direction. Since the enclosed mass, $M_\mathrm{d}$, is an
increasing function of the radial distance, $r$, the effect of
vertical self-gravity becomes progressively important in outer
part of the disc, specially in thick discs (for ADAFs the typical
value of $\frac{H}{r}$ is around 1).

 On the other hand in a self-gravitating disc, the hydrostatic equilibrium equation in the vertical direction yields (e.g. Paczynski 1978, Duschl 2000):
\begin{equation}
P=\pi G \Sigma^{2}
\end{equation}
where $P$ is the pressure in $z=0$ (central plane).
Following (Mosallanezhad et al. 2012)  we assume the disc to be
isothermal in the vertical direction.

In order to complete the problem we need to introduce energy
equation. We assume the generated energy due to viscous
dissipation into the volume is balanced by the advection cooling
and energy loss of outflow. Thus,
\begin{displaymath}
\frac{\Sigma V_r}{\gamma-1}\frac{dc_\mathrm{s}^{2}}{dr}-2H V_r
c_\mathrm{s}^{2}\frac{d\rho}{dr}= f \Sigma \nu
r^{2}(\frac{d\Omega}{dr})^{2}
\end{displaymath}
\begin{equation}
-\frac{1}{2}\eta \dot{m}_\mathrm{w}(r)V_\mathrm{k}^{2}(r)
\end{equation}
where $\gamma$, $f$ and $\Omega_\mathrm{k}$ are adiabatic index,
the ratio of specific advection parameter and the Keplerian
angular speed respectively. The last term on the right hand side
of the energy equation represents the energy loss due to wind or
outflow (Knigge 1999). In our model $\eta$ is a free and
dimensionless parameter. The large $\eta$ corresponds to more
energy extraction from the disc because of wind (Knigge 1999).
Finally since we consider the toroidal configuration magnetic
field, the induction equation can be written as:
\begin{equation}
\frac{d}{dr}(V_r B_\varphi)=\dot{B}_\varphi
\end{equation}
where $\dot{B}_\varphi$  is the field escaping/creating rate due
to magnetic instability or dynamo effect.

\section{Self-Similar Solutions}

In the last section we introduced the basic equations for a
vertically self-gravitating, axi-symmetric, magnetized hot
accretion flow in the presence of rotating wind. The basic
equations of the model are a set of partial differential
equations, which have a very complicated structure. The
self-similar method is one of the most useful and powerful
techniques to give an approximate solutions for differential MHD
equations and has a wide range of applications in astrophysics.
For the first time this technique was applied by Narayan \& Yi
(1994) in order to solve ADAFs dynamical equations. By adopting
Narayan \& Yi (1994) self-similar scaling, in fact, the radial
dependencies of all physical quantities are canceled out, and all
of differential equations are transformed to algebraic equations.
Using self-similar scaling and Including the effect of mass
outflow and making the reasonable assumption as introduced by
(Abbassi et al. 2010, Mosallanezhad et al. 2012) the velocities
are supposed to be expressed as follows,

\begin{equation}
V_r(r)=-c_1 \alpha V_\mathrm{k}(r)
\end{equation}
\begin{equation}
V_\varphi(r)=c_2 V_\mathrm{k}(r)
\end{equation}
\begin{equation}
c_\mathrm{s}^{2}=c_3 V_\mathrm{k}^{2}
\end{equation}
\begin{equation}
c_\mathrm{A}^{2} = 2 \beta c_\mathrm{s}^{2}
\end{equation}
where
\begin{equation}
V_\mathrm{k}(r)=\sqrt{\frac{GM}{r}}
\end{equation}

Generally, $c_1$, $c_2$ and $c_3$ possess a radial dependencies. But we will show in Fig. 1 that their slopes with respect to radius are small over the range of radii considered ($r > 30 r_S$) . The deviations will be small as well, and that the strongest deviations occur at the smallest radii. As It is clear c1, c2 and c3 are almost constant in the range of $r = 30 r_s$ to $r = 100 r_s$. This is the most accepted range for validity of advection dominated disks.
On the other hand the self-similar solutions that we had been adopted are valid far from the boundaries. It is clear that in the main body
of the disk where the self-similar assumption is valid, $c_i$ are almost constant. Hereafter we used constant $c_1$, $c_2$ and $c_3$ and they are determined later from the
basic equations. Assuming the surface density $\Sigma$ to be in
the form of:
\begin{equation}
\Sigma=\Sigma_0 r^{s}
\end{equation}
where $s$ is constant. In order to have a valid solution for the
self-similar treatment, the mass-loss rate per unit volume and
the field escaping rate must have the following form:
\begin{equation}
\dot{\rho}=\dot{\rho_0} r^{2s-\frac{1}{2}}
\end{equation}
\begin{equation}
\dot{B_\varphi}=\dot{B_0} r^{s-\frac{3}{2}}
\end{equation}

\input{epsf}\epsfxsize=3.8in \epsfysize=3.8in\begin{figure}\centerline{\epsffile{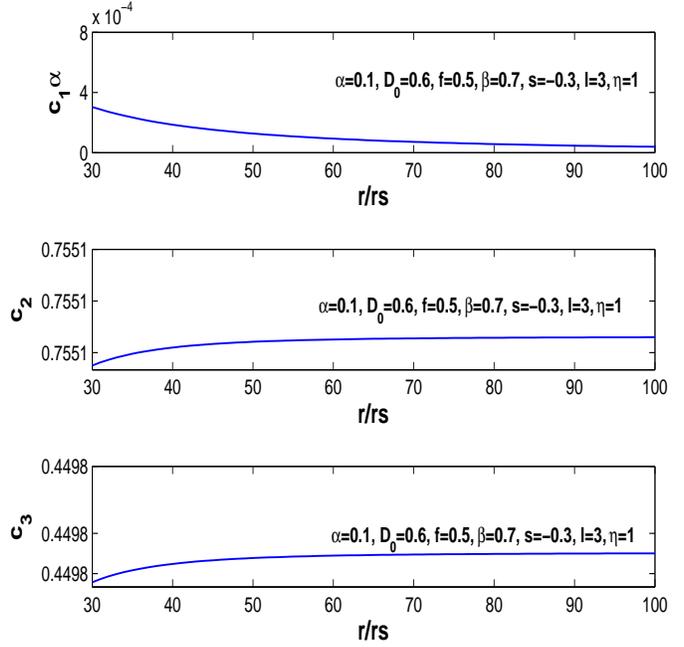}}
\caption{ Numerical coefficient $c_i$ as function of radius. This range is most accepted radial range for ADAFs. 
 for a given values of input parameters.}
\end{figure}

Considering hydrostatic equation, we obtain the disc
half-thickness $H$ as:
\begin{displaymath}
H=\frac{c_{3}(1+\beta)}{2\frac{M_\mathrm{d}}{M_{\star}}}
r=\frac{c_{3}(1+\beta)}{D} r
\end{displaymath}
$D(=2\frac{M_\mathrm{d}}{M_{\star}})$ is dimensionless parameter
and  represents the importance of vertical self-gravitation of
the disc. By substituting the above self-similar solutions in to
the dynamical equations of the system, we obtain the following
system of dimensionless equations, to be solved for $c_1$, $c_2$
and $c_3$:
\begin{equation}
\dot{\rho_0}=-(s+\frac{1}{2})\frac{c_1 \alpha \pi (\Sigma_0)^{2}
\sqrt{GM_{\star}}}{2c_{3}(1+\beta)M_{\star}}
\end{equation}
\begin{equation}
H=\frac{c_{3}(1+\beta)M_{\star}}{2\pi\Sigma_0}r^{-s-1}
\end{equation}

By using equations (1),(2) and (3) we have:
\begin{equation}
\dot{M}_\mathrm{w}=\dot{M}_\mathrm{0w}r^{s+\frac{1}{2}}
\end{equation}
where
\begin{equation}
 \dot{M}_\mathrm{0w} = 2 \pi \alpha c_1 \sqrt{G M_{\star}} \Sigma_0
\end{equation}
\begin{equation}
\dot{m}_\mathrm{w} = \frac{s + \frac{1}{2}}{2} \alpha c_1 \sqrt{G
M_{\star}} \Sigma_0 r^{s - 3/2}
\end{equation}
we define $\dot{m}$ as:
\begin{equation}
\dot{m}=\frac{\dot{M}_\mathrm{0w}}{\pi\alpha \Sigma_{0}\sqrt{GM_{\star}}}
\end{equation}
so
\begin{equation}
\dot{m}=2c_{1}
\end{equation}
equations (28-31) come from equations (4),(5),(11) and (12)
respectively:
\begin{equation}
-\frac{1}{2}c_1^{2}\alpha^{2}=c_2^{2}-1-[s-1+\beta(s+1)]c_3
\end{equation}

\begin{equation}
c_1=\frac{-3(1+\beta)}{D}c_{3}^{\frac{3}{2}}+\dot{m}(s+\frac{1}{2})l^{2}
\end{equation}
\begin{equation}
[\frac{1}{\gamma-1}+(2s+1)]c_1 c_3 =\frac{9}{4 D}f
c_3^{\frac{3}{2}} c_2^{2}(1+\beta)-\frac{1}{8}\eta
(s+\frac{1}{2})\dot{m}
\end{equation}
\begin{equation}
\dot{B}_{0}=\frac{(1-2s)}{2}\alpha\Sigma_{0} c_{1} G \sqrt{M_{\star}}\sqrt{\frac{2
 \beta}{1+\beta}}
\end{equation}

As it is easily seen from equation (21), for $s = -1/2$, there is
no mass loss/wind, while there exists mass loss (wind) for $s >
-1/2$.  On the other hand, the ecape/creation of magnetic fields
will balance each other  \textbf{for$s= \frac{1}{2}$} (equation 31). In this work we
focus on the wind \textbf{case($-\frac{1}{2}<s<\frac{1}{2}$)}. Furthermore the
thickness of the disc will increase due to the magnetic pressure
for the weakly to moderately magnetized flow with $ \beta \sim
1$; while it decreases when D is relatively large.

After algebraic manipulations, we obtain a sixth order algebraic
equation for $c_1$:
\begin{equation}
A^{3}c_1^{6}-3(1-E)A^{2}c_1^{4}+[B^{3}+3A(1-E)^{2}]c_1^{2}-(1-E)^{3}=0
\end{equation}
Where
\begin{equation}
A=\frac{1}{2}\alpha^{2}
\end{equation}
\begin{displaymath}
B=\frac{4}{3^{\frac{5}{3}}f}[\frac{1}{\gamma-1}+(2s+1)](\frac{D}{1+\beta})^\frac{2}{3}[\frac{1}{-1+2(s+\frac{1}{2})l^{2}}]^{\frac{1}{3}}
\end{displaymath}
\begin{equation}
-[s-1+\beta(s+1)][\frac{D}{3(1+\beta)}(-1+2(s+\frac{1}{2})l^{2})]^{\frac{2}{3}}
\end{equation}
\begin{equation}
E=\frac{\eta}{3f}[\frac{s+\frac{1}{2}}{-1+2(s+\frac{1}{2})l^{2}}]
\end{equation}

Mosallanezhad et al.(2012) solved the equation when
$s=-\frac{1}{2}$ because they did not consider wind/mass-loss in
their model. But we are interested in analyzing the dynamical
behavior of magnetized ADAFs in presence of mass-loss. This
algebraic equation shows that the variable $c_1$ which determines
the behavior of radial velocity depends only on the $\alpha$,
$s$, $D$, $\beta$ and $f$. As we can see in above
equations, in order to have the real solutions for $c_{1}$, we
need to have $l$ as $l^{2}>\frac{1}{2s+1}$. This new requirement limits the solutions of $c_3>0$. Using $c_1$ from this algebraic equation, the other
variables (i.e. $c_2$ and $c_3$) can be obtained easily:
\begin{equation}
c_2^{2}=\frac{4 D c_3^{-\frac{1}{2}}c_1}{9
f(1+\beta)}[\frac{1}{\gamma-1}+(2s+1)] +\frac{\eta (s+\frac{1}{2})
D c_1 c_3^{-\frac{3}{2}}}{9 f(1+\beta)}
\end{equation}
\begin{equation}
c_3=c_1^{\frac{2}{3}}[\frac{D
(-1+2(s+\frac{1}{2})l^{2})}{3(1+\beta)}]^{\frac{2}{3}}
\end{equation}

We can solve these simple equations numerically, and clearlyjust physical solutions can be interpreted. They reduce to the resultsof Mosallanezhad et al. (2012) without wind. Now we can analyze thebehavior of solutions.

\section{Results}
Now, we have performed a parameter study considering
our input parameters. We are interested to examine the effects of
rotating wind, magnetic field, self-gravity and advection which their
characteristic parameters are :$s$,$l$, $\beta$, $D$ and
$f$, respectively. We solved the equations of $c_i$
numerically. According to the new condition for $l$ we use
$l>1$ to get real values for $c_1$, $c_2$ and $c_3$. $l>1$ corresponds to rotating wind which can remove a
significant amounts of angular momentum from the disc. Our results for the
structure of vertically self-gravitating hot magnetized ADAFs are
shown in Figures 2-6. In all these figures, the necessary
constant are fixed to their most typical values. In all
figures we use $\gamma=\frac{5}{3}$. Fig. 2 shows the
coefficients $c_1, c_2$ and $\frac{H}{r}$ in terms of
self-gravitating parameter $D$ for different values of
dimensionless lever arm $l$ ($l=2.,2.5,3.$). In Fig. 2 we
investigate the role of the extraction of angular momentum due to
the wind/outflow in the radial velocity, rotation velocity and the
thickness of accretion flow.  The behavior of radial velocity is
determined by $c_1 \alpha$ is shown in the upper panel. Top
panel in Figure 2 shows that for larger values of parameter $D$,
which implies self-gravitating becomes important, the radial
velocity decreases sharply as parameter $D$ increases.
Furthermore when $l$ becomes stronger the radial velocity steadily
increases which makes sense and is in great agreement with
Abbassi et al (2013). The rotational velocity, $c_2$, is plotted
in the middle panel. As we can see easily the rotational velocity
is almost independent of the self-gravity's parameter $D$.
Abbassi et al (2013) have shown that the accreting flow will
rotate slower when the angular momentum removal from the disc, as
$l$ becomes larger. This trend is observed in our results (middle
panel of Figure 2). Bottom panel of Figure 2 show the relative
thickness of disc, $H/r$ as a function of $D$ for different values
of $l$ parameter. For instance, one would immediately deduce that
the relative thickness decreases meaningfully at the outer part
of the disc where the disc becomes self-gravitating (larger
values of $D$). On the other hand for non-self-gravitating discs,
small $D$, the half-thickness increases when the amount of the
extracted angular momentum becomes stronger. These result agree
with previous solution (Abbassi et al. 2013). In Figure 2 each
curve is labeled by corresponding index $l$.

Fig. 3 shows the coefficients $c_1, c_2$ and $\frac{H}{r}$
in terms of self-gravitating parameter $D$ for different values
of wind parameter $s$. For lower values of $D$,
non-self-gravitating case, the radial velocity is increases when
the wind becomes stronger, which is in agreement with Abbassi et al (2013).  The
rotational velocity is almost independent of the self-gravity's
parameter $D$, Middle panel. We increase the wind parameter, we see rotational velocity decreases. The relative thickness of disc, $H/r$ decreases gradually at the outer part of the disc where the disc becomes self-gravitating (larger values
of $D$). Also for non-self-gravitating discs, Lower $D$, the
half-thickness increases when the wind is stronger ($s>-1/2$).
These result agree with previous solutions (Abbassi et al. 2013,
Mosallanezhad et al. 2012). In Figure 2 Each curve is labeled by
corresponding index $s$.

\input{epsf}\epsfxsize=3.8in \epsfysize=3.8in\begin{figure}\centerline{\epsffile{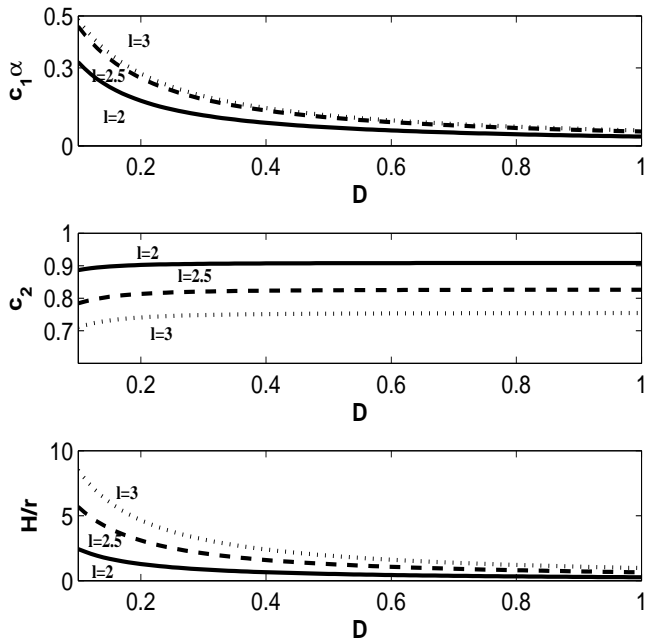}}
\caption{ Numerical coefficient $c_i$ as function of self-gravity
parameter $D$ for several values of $l$ (the amount of the
extracted angular momentum). For all panels we use
$s=-0.3$, $\beta=0.7$, $\alpha=0.1$, $\eta=1.$, $f=0.5$. }
\end{figure}

\input{epsf}\epsfxsize=3.8in \epsfysize=3.8in\begin{figure}\centerline{\epsffile{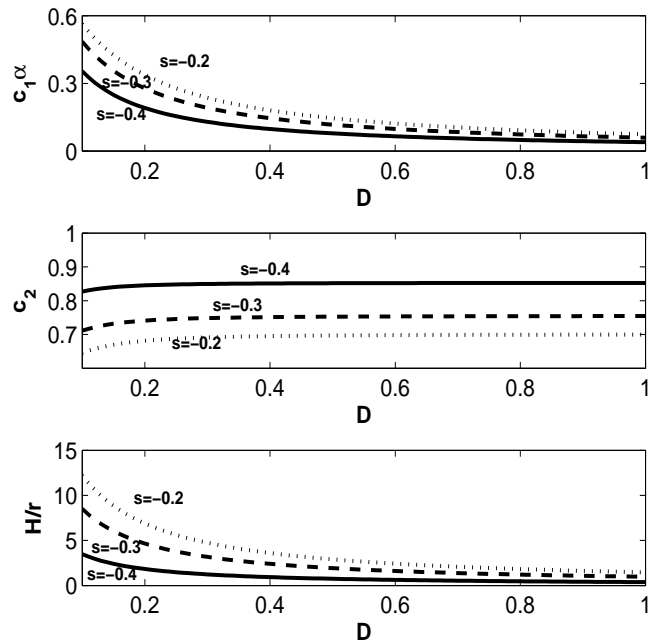}}
\caption{ Numerical coefficient $c_i$ as function of self-gravity
parameter $D$ for several values of $s$ (wind parameter).
For all panels we use $l=3.$, $\beta=0.7$, $\alpha=0.1$,
$\eta=1.$, $f=0.5$. }
\end{figure}

In Figure 4.  the behavior of the  coefficients c1 and c2
 and relative thickness  are shown for different values of
advective parameter $f$ versus parameter $D$. As disc becomes
advective, larger $f$, the radial velocity and half-thickness
will increase relatively, while the rotational velocity
decreases, particularly for smaller values of $D$. The rotational
velocity decreases when advection parameter $f$ increases and is
not affected by the self-gravity parameter. But radial velocity
and vertical thickness are sensitive to $D$ parameter and
decrease as $D$ increases.

Similar to Figure 2, 3 and 4 the coefficients c1 and c2
and relative thickness are shown in Figure 5 for different values
of magnetic parameter $\beta$ versus parameter $D$. Generally when
the magnetic field becomes stronger, when the $\beta$ increases,
the flow rotate faster, with much more radial velocity and larger
half-thickness compare to non-magnetized case.  It would be
interesting if we study the influence of the input parameters on
the temperature gradient of the disc. Finally, in Figure 6 we
have shown the influence of the wind (upper panel),
self-gravity(middle panel) and the magnetic field(lower panel) on
the radial temperature structure of the flow . As we know, ADAFs
occur in two regime depending on their mass accretion rate and
optical depth. In the limit of low mass accretion rates, we have
optically thin discs. In optically thin ADAFs, the cooling time
of accretion flow is longer than the accretion time scale. The
generated heat by viscosity remains mostly in the accretion disc.
The disc can not radiate their energy efficiently. In this model,
we may estimate the isothermal sound speed as (Akizuki \& Fukue
2006):
\begin{equation}
\frac{R}{\bar{\mu}}T=c_\mathrm{s}^{2}=c_3\frac{GM_{\star}}{r}
\end{equation}
 where $T$ is the gas temperature, $R$ the gas constant and
 $\bar{\mu}$ the mean molecular weight
($\bar{\mu}=0.5$). So,
 the temperature is expressed as:
 \begin{equation}
T=c_3\frac{c^{2}\bar{\mu}}{2R}(\frac{r}{r_\mathrm{s}})^{-1}=2.706 \times10^{12}c_3(\frac{r}{r_\mathrm{s}})^{-1}
\end{equation}
where $r_\mathrm{s}(=2\frac{GM_{\star}}{c^{2}})$ and $c$ are the
Schwarzschild radius of the central object and light speed
respectively. In this formula the coefficient $c_3$ implicitly
depends on the wind, self-gravitation, magnetic field and
advection parameters, ($s, D, \beta, f$). In Fig. 5, we show the
radial behavior of temperature for different value of $s, D$ and
$\beta$. It is obvious that the surface temperature decreases
monotonically as $\frac{r}{r_\mathrm{s}}$ increases. As we can see in the
top panel of Fig. 6, for the case of strong wind ($s>-0.5$) the
surface temperature of the disc will increase
significantly, at least in the inner part of the disc, and this
can impact the observed spectrum since most observed luminosity
of ADAFs comes from inner most region.

 In the lower panel of Fig. 6., we will see the effect of the magnetic field parameter, $\beta$, on the surface
temperature of disc. The surface temperature increases by
increasing $\beta$. Finally in the lower panel we have plotted the surface temperature for
several values of the self gravity parameter $D$.
Temperature gradient of the disc is not very sensitive to the
value of $D$, while though $\beta$ and $s$ more
significant effect are observed. Also it is not easy to calculate the
radiative spectrum of optically thin ADAFs. This model of ADAFs
do not radiate away like a black body radiation. Since accreting
gas in a hot accretion flow has a very high temperature and is
moreover optically thin and magnetized, the relevant radiation
processes are synchrotron emission and bremsstrahlung, modified by
Comptonization (Yuan et al (2014)). In the other limit, the
optically thick ADAFs or Slim disc, the mass accretion rate and
the optical depth is very high. So the radiation generated by
accretion disc can be trapped within the disc. In optically thick
ADAFs, the radiation pressure dominates and sound speed is
related to radiation pressure. This model radiates away locally
like a black body radiation. The averaged flux $F$ is:
\begin{equation}
\Pi=\Pi_{rad}=\frac{1}{3}a T_\mathrm{c}^{4}2H=\frac{8H}{3c}\sigma
T_\mathrm{c}^{4}
\end{equation}

\begin{equation}
F=\sigma
T_\mathrm{c}^{4}=\frac{3c}{8H}\Pi=\frac{3}{8}c\Sigma_{0}\frac{D}{1+\beta}G
M r^{s-2},
\end{equation}

where $\Pi$, $T_\mathrm{c}$, $\sigma$ is the height-integrated
gas pressure, the disc central temperature and the
Stefan-Boltzman constant respectively. The optical thickness of
the disc in the vertical direction is:

\begin{equation}
\tau=\frac{1}{2}\kappa \Sigma=\frac{1}{2}\kappa \Sigma_{0}r^{s}
\end{equation}
where $\kappa$ is the electron-scattering opacity. So,
the effective temperature of the disc surface become:

\begin{equation}
\sigma T_\mathrm{eff}^{4}=\frac{\sigma
T_\mathrm{c}^{4}}{\tau}=\frac{3c}{4\kappa}\frac{D}{1+\beta}\frac{G
M}{r^{2}}=\frac{3}{4}\frac{D}{1+\beta}\frac{L_\mathrm{E}}{4\pi r^{2}}
\end{equation}

\begin{equation}
T_\mathrm{eff}=[\frac{3L_\mathrm{E}}{16\pi
\sigma}\frac{D}{(1+\beta)}]^{\frac{1}{4}}r^{\frac{-1}{2}}
\end{equation}
where $L_\mathrm{E}=4\pi c \frac{G M}{\kappa}$ is the Eddington
luminosity. As we can see, there is no $f$ and $s$ dependence in
the surface temperature of the optically thick ADAFs. The surface
temperature is influenced by the self-gravity and magnetic field
parameters explicitly. It is clear that the temperature increases
by adding $D$ parameter and decreases for larger values of the
magnetic field ($\beta$). In the self-gravitating optically thick
ADAFs, the surface temperature is not affected by wind parameter.

\input{epsf}\epsfxsize=3.8in \epsfysize=3.8in\begin{figure}\centerline{\epsffile{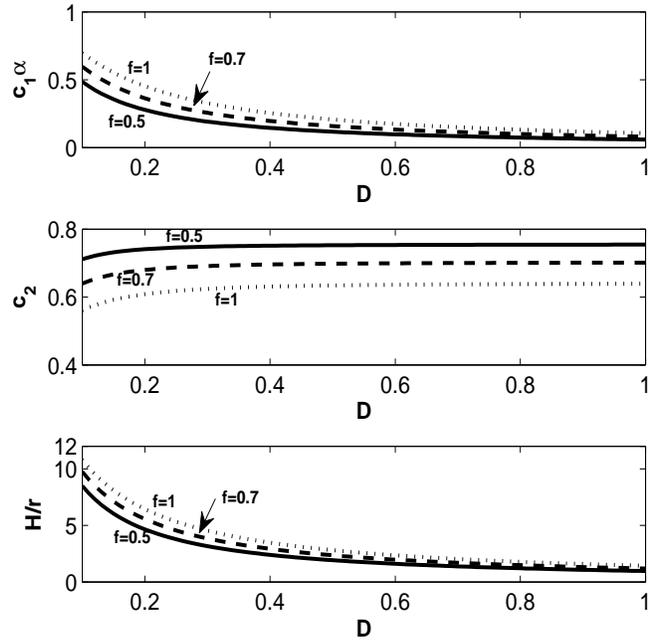}}
\caption{ Numerical coefficient $c_i$ as function of self-gravity
parameter $D$ for several values of $f$ (advection
parameter).For all panels we use $s=-0.3$, $\beta=0.7$,
$\alpha=0.1$, $\eta=1.$, $l=3.$. }
\end{figure}

\input{epsf}\epsfxsize=3.8in \epsfysize=3.8in\begin{figure}\centerline{\epsffile{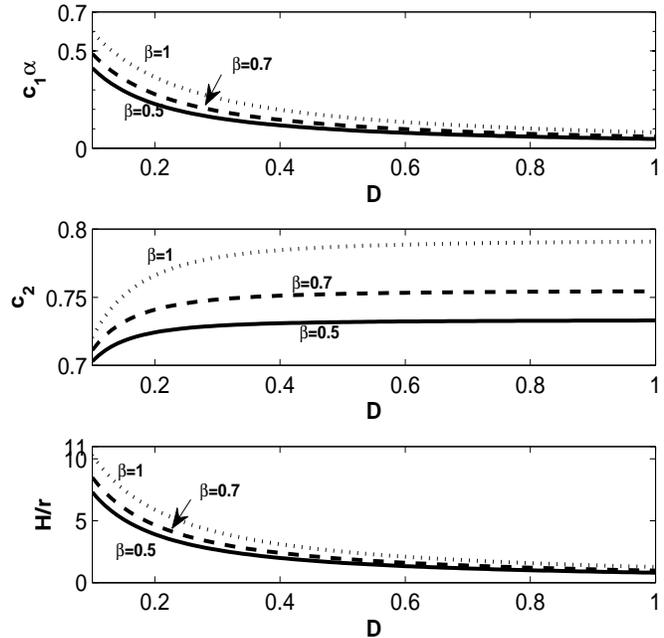}}
\caption{ Numerical coefficient $c_i$ as function of self-gravity
parameter $D$ for several values of $\beta$ (magnetic field
parameter).For all panels we use $s=-0.3$, $f=0.5$,
$\alpha=0.1$, $\eta=1.$, $l=3.$. }
\end{figure}

\input{epsf}\epsfxsize=3.8in \epsfysize=3.8in\begin{figure}\centerline{\epsffile{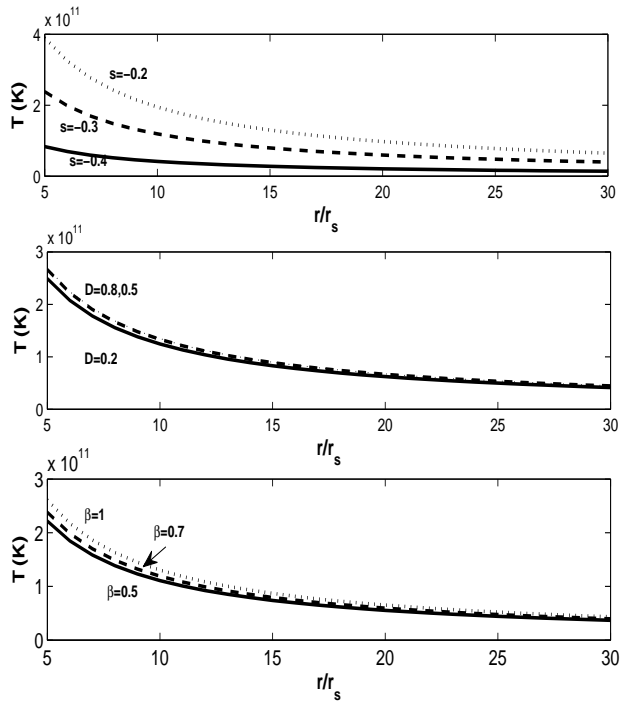}}
\caption{ This plot shows The surface temperature of the disc
$T(k)$ as function of dimensionless radius ($\frac{r}{r_\mathrm{s}}$)
for several values of (top panel: wind parameter $(s)$, middle
panel: self-gravity parameter $D$ and bottom panel: magnetic
field parameter ($\beta$).Numerical coefficient $c_i$ as function
of self-gravity parameter $D$ for several values of $f$ (advection
parameter).For all panels we use $\eta=1$.}
\end{figure}

\section{Summary and Conclusion}
In this paper, we have studied the accretion disc around black
hole in an advection dominated regime in the presence of a
toroidal magnetic field and vertical self-gravity of the disc. It
was assumed that disc wind/outflow contributes to loss of mass,
angular momentum, and thermal energy from accretion discs. We
used the self-similar method for solving the equations. Although
the self-similar solutions are too simple, they improve our
understanding of the physics of the accretion discs around black
hole. For simplicity, we assume an axially symmetric and static
flow with $\alpha$ prescription of viscosity. Also we ignore the
relativistic effects and we use newtonian gravity in the radial
direction. We consider the vertical self-gravity of disc by
following the paper of Mosallanezhad et al. (2012) in the
presence of the effect of wind. Our results reproduce their
solutions when the effect of wind is neglected. In ADAFs, the
more dissipated energy is advected in the flow and so ADAFs are
hot and thick. The disc rotates slower and becomes
thicker in the presence of strong rotating wind. Also we have
shown the self-gravity and wind parameter have the opposite
effects on the thickness and radial velocity of the disc. The
rotational velocity almost is not sensitive to the self-gravity
while it has a more significant effect to the wind parameter.
Beside we have shown the self-gravity and wind parameter
have the same effects on the surface temperature in the optically
thin ADAFs.

The authors thank the anonymous referee for the careful reading
of the manuscript and his/her insightful and constructive
comments. S. Abbassi acknowledges the support from the Abdus Salam International Centre for Theoretical Physics (AS-ICTP) for his visit through the regular associateship scheme

\end{document}